# Operator-in-the-Loop Rolling Dispatch for Damage-Assessment Crews Under Partial Storm Information

Hritik Gopal Shah, Catherine Tajmajer, Elli Ntakou, and Lavelle Freeman
Eversource Energy
Berlin, CT, USA
hritik.shah@eversource.com, catherine.tajmajer@eversource.com, elli.ntakou@eversource.com, lavelle.freeman@eversource.com

**Abstract - Damage assessment (DA) is the first field step between outage identification and repair planning during a major storm. This paper presents an operator-in-the-loop rolling dispatch framework that reads the current outage-management-system (OMS) snapshot, jointly matches available DA crews to visible tickets using road-network travel and operational priority, and returns short pipelines that operators may accept, edit, lock, or reject. The innovation is the deployable OMS-connected decision loop under partial storm information. A chronological replay of one March 7, 2025 storm across three utility area work centers (AWCs) compares the framework with historical business-as-usual (BAU), nearest-neighbor, and customer-weighted policies on a common travel basis. In this single-event replay, aggregate mileage decreased by 36.3%, with AWC-level reductions ranging from 22.9% to 52.1%. Early Assessed Customer Count (EACC) shows how quickly the dispatch sequence exposes damage affecting customers and priority locations to restoration planners; it is a leading assessment metric.**

## I. INTRODUCTION

High-impact weather can damage many distribution assets within a short period and create a restoration workload that changes continuously. Utilities invest in hardening, vegetation management, outage prediction, and pre-staging, but field information remains incomplete after a storm begins [1]-[4]. Before repair activities can be planned, DA crews must locate damaged equipment, identify failure modes, evaluate access and safety constraints, and report the work scope. DA therefore converts an uncertain outage picture into an executable restoration work plan.

Most restoration optimization begins after damaged locations and repair requirements are known. The operational problem considered here occurs earlier: tickets continue to enter OMS, crew positions are represented by their last confirmed locations, and operators need a short plan that can be revised whenever OMS information is updated. At each decision epoch, the proposed tool uses only visible tickets, eligible crews, confirmed anchors, and operator-imposed locks or freezes. A priority-aware matching problem is solved sequentially to form a short pipeline for each crew. The first accepted tasks are frozen after publication, while later tasks remain adaptable.

The contribution is operational rather than a claim of stochastic look-ahead optimality. First, the framework connects directly to OMS and preserves OMS as the system of record. Second, it converts the current event stream into coordinated, operator-reviewable crew pipelines with locks, freezes, and repeated replanning. Third, it evaluates this deployable workflow against historical BAU and two transparent policies using the same chronology and road-network distance basis. The central innovation is the closed loop among OMS state, optimization, operator judgment, and field-status feedback.

## II. RELATED WORK AND SCOPE OF COMPARISON

Utility resilience research spans pre-event prediction and post-event restoration. Outage prediction combines weather, vegetation, infrastructure, and historical data for pre-staging [1]-[4]. Post-disaster models optimize repair routing, reconfiguration, distributed generation, mobile resources, and coordinated restoration [5]-[10]. Recent formulations connect DA with restoration decisions [11], [12], while dynamic vehicle-routing research addresses evolving requests and network states [13]-[14].

Those advanced models answer a broader but different question. They commonly require a validated damage set, repair durations, network topology, switching alternatives, or a probability model for future requests. During initial DA dispatch, those inputs are precisely what crews are collecting, and future OMS tickets are not yet observable. A direct numerical comparison would therefore either give an advanced model future information unavailable to the production tool or require assumptions that cannot be validated from the replay.

The benchmarks were selected to isolate deployable operating choices under identical information: historical BAU, road-distance-only nearest neighbor, and customer-weighted assignment. The proposed policy itself uses optimization and simultaneous maximum-weight bipartite matching but adds rolling OMS ingestion, sequential anchors, stable first tasks, and operator control. Advanced stochastic routing remains an important future benchmark when multiple storms and the required repair and network data become available. The present scope tests whether an OMS-connected policy improves the actual DA decision stage without assuming unavailable future information.

## III. OPERATIONAL CONTEXT AND WORKFLOW

### *A. Business-as-Usual Dispatch*

Storm DA operations are organized by AWC. Tickets enter OMS from customer calls, devices, and operator updates. Each ticket carries a location and operational attributes, while crew status moves among available, assigned, en route, on site, and cleared. Under BAU, an operator scans the visible queue, considers urgency and local geography, selects a crew, communicates one assignment, and waits for the next status update. During heavy inflow, decisions made at different times can conflict, send crews back through covered territory, or delay a high-impact ticket that is not visually prominent.

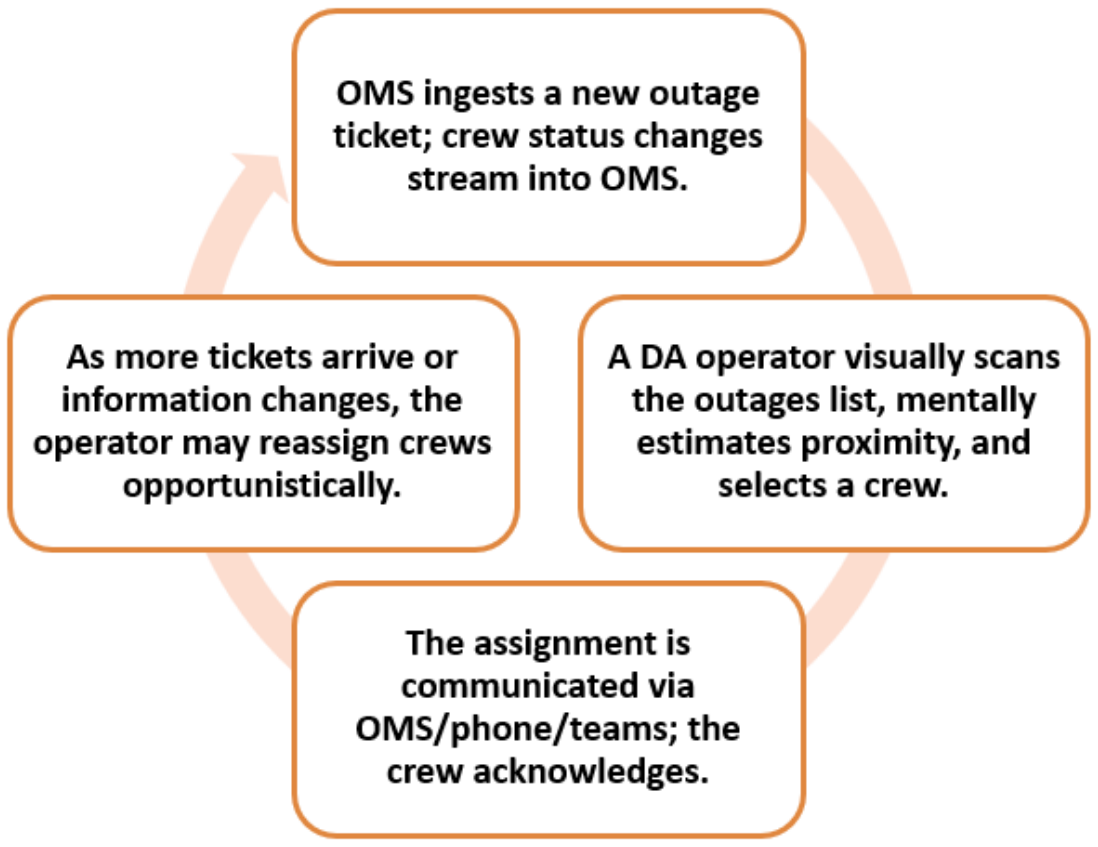

Fig. 1. Business-as-usual damage-assessment dispatch cycle.

### *B. OMS-Connected Rolling Workflow*

The proposed tool is a decision-support layer connected to OMS; OMS remains the system of record. At each fixed or operator-requested trigger, the tool reads visible tickets, crew status, last confirmed locations, and locks or freezes. Candidate tasks are restricted to one AWC, duplicate or infeasible tickets are removed, and road-network travel is computed for feasible crew-ticket pairs.

The allocator recommends an ordered pipeline for each eligible crew. Crews are considered together at each stage, reducing conflicts created by independent assignments. The operator may accept, modify, reject, lock, or freeze any recommendation. After approval, the first task is written to OMS and frozen to avoid field churn. Later tasks remain provisional and may change when tickets or crew states are updated. Fig. 2 shows this closed loop.

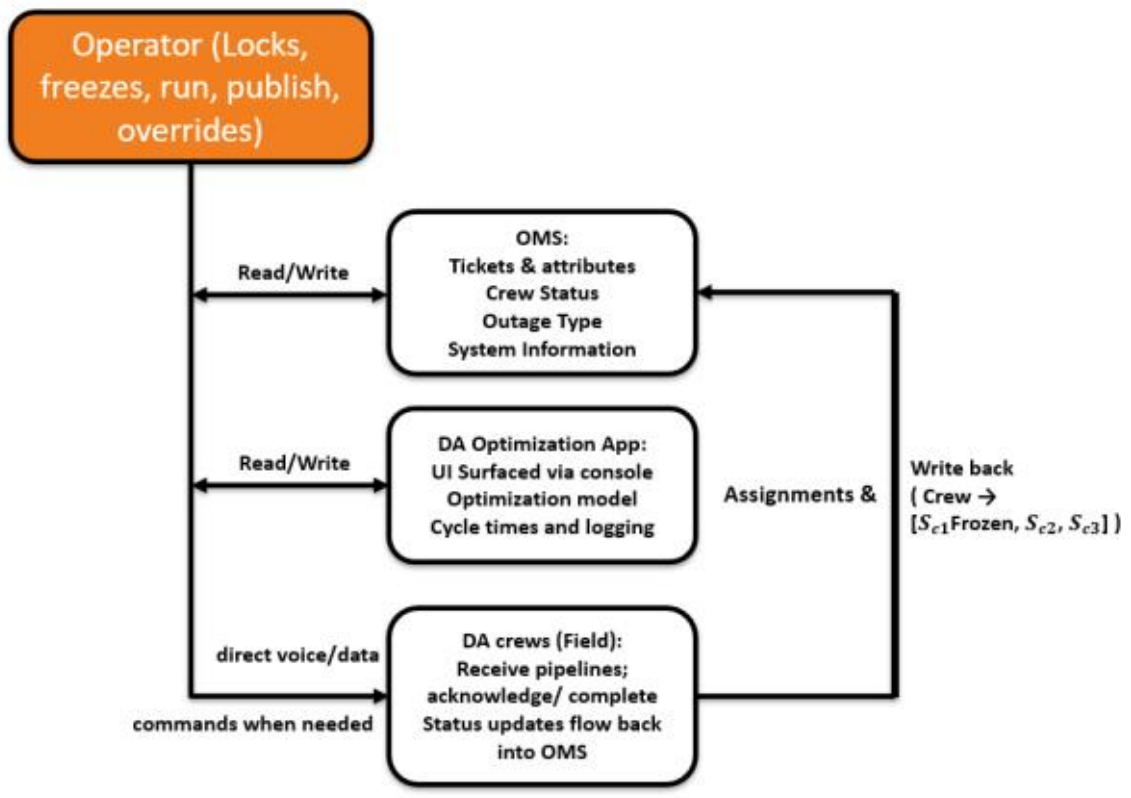

Fig. 2. Operator-in-the-loop rolling DA dispatch workflow.

## IV. ROLLING DISPATCH OPTIMIZATION

### *A. Sets, State, and Priority Value*

For AWC $a$ at decision time $t$, let $C_a$ denote the DA crews and $R_a(t)$ the tickets visible in OMS. $E_c(t) = 1$ indicates that crew $c$ is eligible after status, shift, lock, and freeze rules. The available set is

$$A(t) = \{c \in C_a \mid E_c(t) = 1\} \quad (1)$$

Each crew has an anchor $L_c^k(t)$ equal to its yard at storm start or its last confirmed or assigned DA location at stage $k$. For outage $r$, $\tau_{cr}^k(t)$ is road-network distance from the current anchor.

The OMS priority score $w_r$ follows the utility's deterministic ordering of structured attributes: a priority-customer or critical-location flag is considered first, affected-customer count second, and ticket age third. The resulting ordinal score is expressed in travel-equivalent units. Because multiplying all assignment values by a positive constant does not change the matching, the replay is reported in the normalized form $\beta_{\text{dist}} = 1$:

$$\pi_{cr}^k(t) = w_r - \beta_{\text{dist}} \tau_{cr}^k(t), \qquad \beta_{\text{dist}} = 1 \quad (2)$$

This normalization makes the tradeoff interpretable: increasing $\beta_{\text{dist}}$ approaches nearest-neighbor behavior, whereas decreasing it approaches priority-only assignment. The value changes recommendations, not operator authority; access, safety, and field knowledge can always override the score.

### *B. Sequential Matching and Pipeline Construction*

For pipeline position $k$, $x_{cr}^k(t) = 1$ when crew $c$ is assigned to outage $r$. The stage problem is

$$\max_x \sum_{c \in C_a} \sum_{r \in R_a^k(t)} \pi_{cr}^k(t) x_{cr}^k(t) \quad (3)$$

subject to

$$\sum_{r \in R_a^k(t)} x_{cr}^k(t) \le E_c(t), \qquad c \in C_a \quad (4)$$

$$\sum_{c \in C_a} x_{cr}^k(t) \le 1, \qquad r \in R_a^k(t) \quad (5)$$

After each stage, assigned tickets $A^k(t)$ are removed and crew anchors advance:

$$R_a^{k+1}(t) = R_a^k(t) \setminus A^k(t) \quad (6)$$

$$L_c^{k+1}(t) = L_r, \qquad x_{cr}^k(t) = 1 \quad (7)$$

The problems are solved sequentially, so the method is myopic rather than a joint stochastic look-ahead model.

Repeating the matching $K$ times creates an ordered route without assuming knowledge of tickets arriving later. Locks fix a crew-ticket pairing; a freeze removes a crew or published first task from reconsideration.

### *C. Execution and Baseline Definitions*

At the beginning of a storm, a crew anchor is the AWC yard, or another confirmed starting location. During operation, the anchor is updated to the location of the most recently assigned assessment. The first matching therefore allocates one visible outage to each eligible crew from its current anchor. Selected tickets are then removed, crew anchors are advanced, and the same matching is solved again to generate the next pipeline position. Repeating this process K times creates an ordered route without if tickets arriving later in the storm are already known. The complete pipeline is a recommendation rather than an automatic command. Before publication, the DA operator can accept, modify, reject, lock, or freeze assignments using current field knowledge. A lock preserves a specific crew-ticket pairing. A freeze prevents a crew or published first task from being reconsidered. Once the plan is approved, the first task is written to OMS as the stable field instruction, while later tasks remain provisional and may be recomputed when new tickets arrive or crew status changes. This rolling execution separates two-time scales. Within one decision epoch, the sequential matching coordinates all currently eligible crews and builds short route structure. Across decision epochs, new OMS snapshots provide adaptation to evolving storm information. The approach therefore offers recourse through repeated replanning, but it does not claim stochastic anticipation of future outages. The value of K determines how much downstream outage recommendation is created in each cycle; the measured sensitivity in Section V-D is used to select K=3.

ALGORITHM 1. PROPOSED ROLLING DAMAGE-ASSESSMENT WORKFLOW

| | |
|---|---|
| **Input** | Current OMS snapshot, selected AWC, road network travel matrix, operator locks/freezes, and maximum pipeline length K. |
| **Output** | Ordered pipeline $S_c = [S_{c1} \ldots, S_{ck}]$ for each eligible crew. |
| 1 | Trigger on the fixed cadence or when an operator requests a new allocation. |
| 2 | Read visible tickets, crew availability, last confirmed anchors, and current locks/freezes at time t. |
| 3 | Restrict tickets and crews to AWC a; remove unavailable crews and infeasible or already locked assignments. |
| 4 | For k=1,…,K, compute crew-ticket travel and net values, then solve (3) subject to (4)– (5). |
| 5 | Append selected tickets to each crew pipeline; remove assigned tickets and update anchors (6)– (7). |
| 6 | Present the pipelines to the operator, who may accept, modify, reject, lock, or freeze an assignment. |
| 7 | Publish the approved plan to OMS, freeze the first task, log the snapshot and edits, and repeat at the next trigger. |

## V. CASE STUDY AND RESULTS

### *A. Chronological Replay and Measures*

The March 7, 2025 storm was replayed in three AWCs containing 27 DA crews and 272 tickets. BAU retains the actual OMS crew-assignment sequence. A ticket enters the simulated candidate set only when it appeared in the recorded chronology; no policy receives future ticket information. Every observed or simulated route leg uses the same post-storm road-network distance matrix.

BAU nevertheless contains field frictions that the simulated policies do not reproduce, including access delays, safety holds, informal rerouting, and deviations from the recorded assignment. Simulated policies execute each recommended leg as recorded in the replay. This asymmetry can inflate the apparent mileage savings of every simulated method relative to BAU.

Let $n_r$ be the affected-customer count on ticket $r$ and $S_{c,j}$ crew $c$'s ticket at pipeline position $j$. With duplicate assignment prohibited, EACC at horizon $h$ is

$$\text{EACC}@h = \sum_a \sum_{c \in C_a} \sum_{j=1}^{\min(h,|S_c|)} n_{S_{c,j}} \quad (8)$$

For example, if two crews' first two tickets affect (120, 40) and (80, 20) customers, respectively, EACC@2 = 260. EACC is aggregated across crews and AWCs and quantifies how quickly assessment reaches customer exposure. Combined with the priority-customer rule in the assignment score, it supports earlier discovery of damage affecting large customer groups and priority locations, which can reduce information delay before repair planning. It does not by itself prove shorter restoration time or lower SAIDI/CMI because repair availability, topology, switching, and field duration are not modeled.

### *B. Comparison with BAU and Policies*

TABLE I. AGGREGATE THREE-AWC SINGLE-EVENT RESULTS

| Method | Miles | Red. (%) | EACC @3 | EACC @5 |
|---|---|---|---|---|
| Historical BAU | 1,930 | — | 446 | 523 |
| Nearest neighbor | 1,250 | 35.2 | 527 | 590 |
| Customer-weighted | 1,280 | 33.7 | 759 | 906 |
| Proposed rolling matching | 1,229 | 36.3 | 631 | 713 |

The proposed workflow produced the lowest aggregate mileage, reducing travel from 1,930 to 1,229 miles in this event. Nearest neighbor was close on mileage, confirming that proximity explains much of the benefit, but its EACC values were lower. Customer-weighted dispatch produced the highest EACC because it emphasizes customer count directly; it traveled 51 miles more and does not balance travel and priority in the same score. The proposed method combines low mileage, earlier customer-impact assessment than BAU and nearest neighbor, coordinated crew matching, and operator review. No method dominates every objective.

### *C. Variation Across Work Centers*

TABLE II. PROPOSED SINGLE-EVENT REPLAY RESULTS BY AWC

| AWC | Crews | Tickets | BAU mi | Proposed mi | Red. (%) |
|---|---|---|---|---|---|
| AWC 1 | 7 | 90 | 762 | 365 | 52.1 |
| AWC 2 | 6 | 59 | 146 | 76 | 47.9 |
| AWC 3 | 14 | 123 | 1,022 | 788 | 22.9 |
| All | 27 | 272 | 1,930 | 1,229 | 36.3 |

Mileage decreased in each AWC, but the spread was wide: 22.9–52.1%. Avoidable travel depends on crew count, geographic scale, ticket dispersion, outage density, and the efficiency of historical routes. AWC 3 was the largest operation and showed the smallest percentage reduction. The aggregate saving was 701 miles across 27 crews and 272 tickets; it remains a single-event result.

### *D. Pipeline Length and Runtime*

TABLE III. PIPELINE-LENGTH RUNTIME SENSITIVITY IN AWC 1

| Pipeline length K | Median runtime | Operational interpretation |
|---|---|---|
| 1 | 25 s | Fast, but frequent reruns and no downstream structure |
| 3 | 45 s | Sub-minute latency with a stable short pipeline |
| 5 | 120 s | More look-ahead, but Mult minute delay |

The implementation uses Python and solves each stage as a maximum-weight bipartite assignment. Reported runtimes are end-to-end measurements on a 128 GB RAM Microsoft azure cloud server. The assignment solve is subsecond at the tested scale. Road-network matrix retrieval, batching, and repeated matrix construction across $K$ stages dominate the 25–120 s totals.

$K = 1$ supplied only the next assignment and required more frequent reruns. $K = 5$ created more downstream structure but increased delay and produced a plan more likely to become stale. $K = 3$ required 45 s in AWC 1 and yielded three ordered tasks while keeping the first stable and later tasks revisable. With $K = 3$, median runtimes were 45, 52, and 48 s in AWCs 1–3. The sensitivity supports a short operational horizon; it does not convert the method into a globally optimal multi-period route model.

## VI. OPERATIONAL INTERPRETATION AND LIMITATIONS

The results separate three sources of value. Nearest-neighbor performance shows the benefit of using road-network proximity. Customer-weighted performance shows the benefit of explicitly surfacing high-customer tickets. The proposed method combines both signals while coordinating all eligible crews, retaining a short downstream plan, and allowing the operator to preserve or override field decisions. Its contribution is therefore not that maximum-weight matching is new; it is that a transparent matching model is embedded in a rolling OMS workflow that can be used before a complete damage set or restoration model exists.

The priority-travel coefficient has a direct operational interpretation. With priority expressed in travel-equivalent units, $\beta_{\text{dist}} = 1$ is the reported normalized setting. A larger value makes recommendations more local and converges toward the nearest-neighbor benchmark; a smaller value accepts more travel to reach high-priority or high-customer tickets and converges toward the customer-weighted benchmark. This limiting-case comparison provides a qualitative sensitivity check without claiming that one coefficient is universally optimal. In deployment, the score and coefficient should be reviewed at storm start and held fixed during an event so operator behavior remains predictable.

The OMS connection also changes what can be claimed from the evaluation. A purely offline routing model can optimize a fixed list, but the production problem includes ticket arrivals, crew-state changes, locks, rejected recommendations, and a stable first task. These controls are not represented in standard distance-only benchmarks. Mileage reductions therefore estimate dispatch-policy opportunity under consistent information and distance measurement; they are not guaranteed realized savings.

Finally, EACC is a leading information metric. DA must be completed before repair scope, crew type, material needs, and access constraints can be confirmed. Increasing EACC early in the sequence can make actionable damage information available sooner for more affected customers and priority locations. The downstream response still depends on repair crews, topology, switching, materials, and storm conditions.

## VII. CONCLUSION

This paper presented an OMS-connected, operator-in-the-loop framework for dispatching DA crews while outage information is still arriving. In one March 7, 2025 storm replay across three AWCs, the proposed workflow reduced aggregate road-network mileage by 36.3%, with AWC-level reductions ranging from 22.9% to 52.1%.

EACC identifies how early the dispatch sequence assesses damage affecting customers; together with explicit priority-customer ordering, it can deliver actionable damage information to restoration planners earlier. The principal contribution is the production-oriented connection among live

OMS state, coordinated matching, rolling replanning, operator control, and field feedback. Broader storm validation and comparison with advanced stochastic routing are future work once comparable damage, repair, topology, and runtime data are available.